# What is the Power Stroke Mechanism of Muscle Contraction?

**Josh Baker** [1,*]

[1] University of Nevada School of Medicine; jebaker@unr.edu
* Correspondence: jebaker@unr.edu; Tel.: 1-775-351-4243

**Abstract:** Muscle's power stroke occurs when muscle shortens against a mechanical load, performing irreversible work on its surroundings over time. The mechanism of muscle's power stroke is thermodynamic. Nevertheless, for more than 65 years molecular biophysicists have rejected thermodynamics in favor of molecular mechanisms in which biophysicists make in silico molecules do whatever they want them to do. In these effectively linguistical models, the molecular power stroke mechanism is whatever the biophysicist declares it to be even if it is not, in fact, a power stroke. While for 25 years I have tried, it is impossible to formally disprove linguistical models, and so here I focus on empirical data to ask the simple question: what is the power stroke mechanism of muscle contraction? I describe the 65-year search for a molecular power stroke mechanism that, as predicted by A.V. Hill, revealed that a molecular power stroke mechanism does not exist. The primary molecular mechanism observed is a force-generating switch. Meanwhile, after 85 years, A.V. Hill's thermodynamic power stroke mechanism remains unchallenged, and his thermodynamic model, which can be derived from the energetics and mechanics of an ensemble of force-generating switches, remains the most widely used model in muscle physiology. One does not need to understand thermodynamics to empirically conclude that the mechanism of muscle contraction is an irreversible ensemble power stroke, not a reversible molecular power stroke.

## 1. Introduction

Force generation in muscle occurs when myosin motor proteins upon binding to an actin filament undergo a switch-like conformational change that stretches spring-like elements in muscle proteins. Muscle shortening occurs when actin thin filaments slide past myosin thick filaments pulling the ends of a muscle sarcomere (the fundamental contractile unit) closer together. The energy for muscle force generation and shortening is derived from the free energy for adenosine triphosphate (ATP) hydrolysis; however, the mechanism by which this macroscopic chemical energy is converted into mechanical energy remains unclear.

The power stroke of muscle is unambiguously defined as the irreversible work performed with shortening muscle over time. The shortening stroke of muscle is characterized by the distance a half sarcomere shortens, $x$. The work performed by this stroke against a force, $F$, is $Fx$. Muscle's power stroke is characterized by the work, $Fx$, performed with its shortening stroke over time, which can be written $FV$, where $V = \mathrm{d}x/\mathrm{d}t$ is the rate of shortening of a muscle half sarcomere (1).

Muscle force generation is a fundamentally different physical process from that of muscle's power stroke. For example, when we lift a book from a table, our muscles must first generate force equal and opposite the weight of the book before generating power output in lifting the book. Force generation involves work performed within muscle (i.e., energy is gained by muscle in the form of an internal mechanical potential energy), whereas power output involves work performed by muscle on the book (i.e., energy is lost from muscle and gained by the book in the form of gravitational potential energy). Power output is arguably the most important of muscle's physiological functions, which is to say that before we can claim to understand mechanisms of disease, drug action, and regulation on muscle function, we must first establish the mechanism of muscle's power stroke.

In 1938, based on detailed measurements of muscle mechanics and energetics, A.V. Hill demonstrated a simple thermodynamic relationship between muscle force, *F*, muscle power, *FV*, and heat output, establishing muscle's power stroke as a thermodynamic (ensemble) mechanism (2). However, when myosin motors were first observed (3) and isolated as functional chemical molecules (4, 5), molecular biologists began to speculate about the molecular mechanisms that irreversibly contract muscle.

In 1957, electron micrographs of myosin crossbridges (i.e., motors) in muscle inspired A.F. Huxley to propose a molecular power stroke mechanism (6). Huxley reasoned that if the molecules responsible for muscle's irreversible power stroke are observable, then power strokes (Fig. 2A or Fig. 2B) should be observable as structural changes within those molecules. For the next 65 years a scientific community focused their efforts on searching for these structural changes and, as predicted by A.V. Hill, never found them.

In 1998, we observed in muscle a discrete force-generating myosin lever arm rotation induced by actin binding and gated by inorganic phosphate release (7). This force-generating switch is a reversible chemical (binding) step in the actin-myosin ATPase reaction, which raises the question: how can a reversible chemical step that satisfies detailed balance account for muscle's irreversible power stroke?

These discoveries and the questions they pose parallel those made 100 years prior in early studies of heat engines. A.V. Hill's description of muscle as a thermodynamic machine mirrors that of S. Carnot's 1824 description of a heat engine. The assumption that muscle mechanics is defined by classical mechanic descriptions of the reversible chemical kinetics of motor proteins mirrors the classical mechanic kinetic theory of gases formalized by Clausius in 1857. The question as to how reversible chemical kinetics accounts for muscle's irreversible power stroke is Loschmidt's paradox applied to chemical kinetics. While not evident in gases, muscle provides a solution to this paradox.

Penrose writes, "there is no remotely practical procedure for calculating the behavior of a macroscopic body from a detailed calculation of the motions of all of its [constituents]. (8)" While this is true of most macroscopic bodies, it is not true of muscle. In muscle, not only can we calculate; we can observe and correlate the relationship between the behavior of the macroscopic body and the kinetics of its constituent molecular switches, and in 1999, we did just this (9). Mirroring the Maxwell-Boltzmann distribution, we observed a probability distribution of switches that is determined by the ensemble entropy of switches. Mirroring Boltzmann's equation (8, 10), we observed that a time-dependent probability distribution of switches accounts for muscle's irreversible power stroke.

However, while Boltzmann assumed that ensemble entropy exists within molecules, we observe that ensemble entropy describes probabilities themselves. Specifically, the ensemble entropy is a component of the reaction free energy landscape of the ensemble of molecular switches, and a time-dependent change in entropy irreversibly changes the kinetic and energetic constants of reversible chemical switches, causing a re-distribution (a net flux) of reversible switches. Thus, a net flux of molecular switches does not create the irreversible agency of a power stroke; the irreversible change in ensemble entropy causes the net flux. This solves Loschmidt's paradox (reversible chemical switches do not cause muscle's irreversible power stroke). It also demonstrates that molecular mechanisms of irreversible ensemble processes are not possible (molecules cannot change an energy landscape), solving the paradox of Maxwell's demon (a demon with molecular agency cannot change ensemble entropy). It also solves Gibbs' paradox (11), and it solves the paradox of Schrodinger's cat, demonstrating that predictive probabilities, not molecules, are entangled across scales (Fig. 1) (12).

The paradigm in molecular biology that reversible molecular mechanisms describe irreversible biological function is obviously false. Thus, over the past 25 years we have shown that the molecular basis for A.V. Hill's irreversible chemical thermodynamic model of muscle contraction is an ensemble of force-generating switches and demonstrated that this model accounts for most key aspects of steady state and transient muscle mechanics contraction (13–17). Nevertheless, molecular biophysicists reject formal thermodynamic constraints in favor of the unconstrained authority to create whatever molecular power stroke mechanisms they choose. These models are considered useful because one need only imagine a new molecular mechanism, and it can be mathematically incorporated into a molecular power stroke model to account for the latest observations and discoveries. While these exercises in computational quilting may be satisfying, muscle's power stroke is formally constrained by thermodynamics, and it is empirically constrained by 65 years of structural and mechanical studies. Indeed, models of muscle contraction that ignore observed mechanical and structural constraints undermine the fundamental purpose of more than half a century of experimental studies. Here I focus on the simple question: what is the observed power stroke mechanism of muscle contraction?

I describe how over the past 65 years many molecular mechanisms have been proposed and disproven before arriving at a consensus today that the molecular mechanism of muscle contraction is a force-generating switch and that, as predicted by A.V. Hill, a molecular power stroke mechanism does not exist. Meanwhile, after 85 years, A.V. Hill's thermodynamic power stroke mechanism remains unchallenged, and his thermodynamic model, which can be derived from the mechanics and energetics of an ensemble of force-generating switches, remains the most widely used model in muscle physiology. In short, structural and mechanical studies demonstrate empirically what thermodynamics requires formally: muscle's irreversible power stroke is defined by an ensemble of molecular switches, and a molecular power stroke mechanism is not possible.

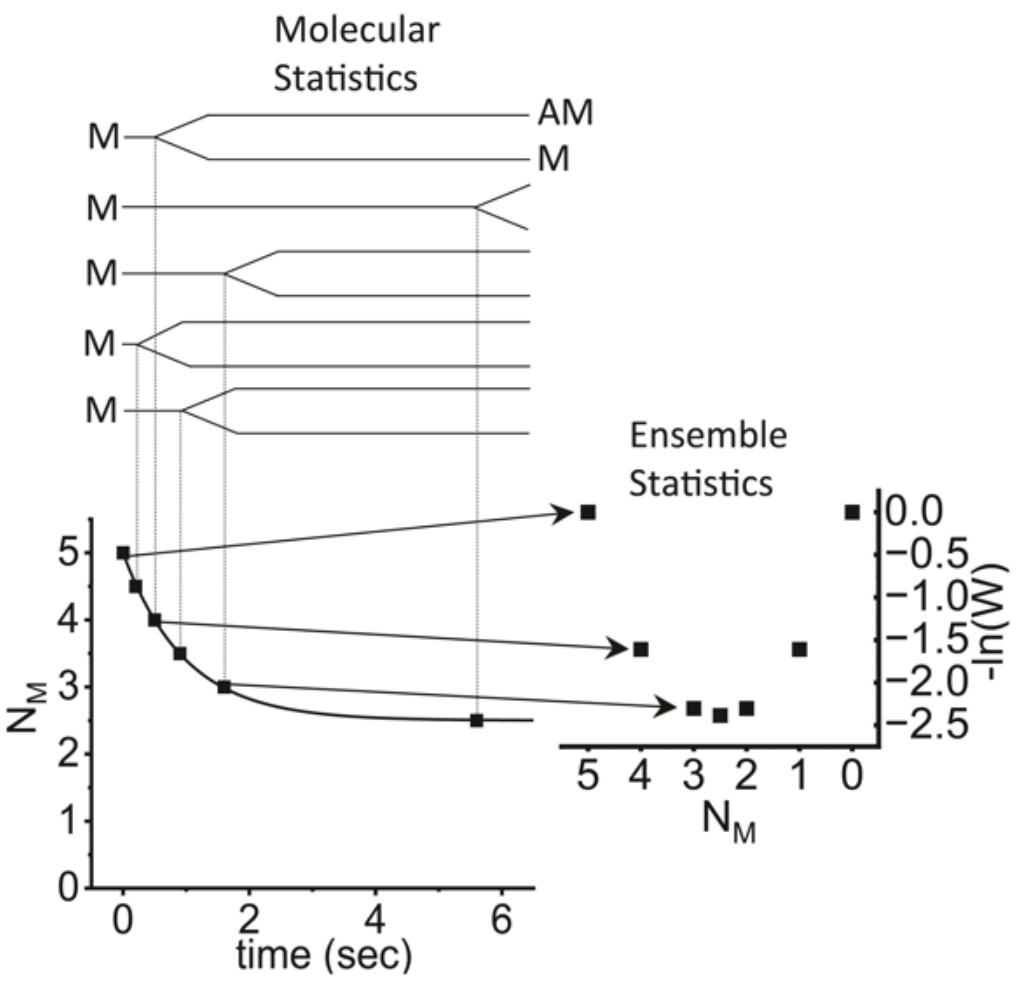

**Figure 1.** Molecular and ensemble kinetics of five molecular switches. The predictive statistics (Molecular Statistics, upper left) of five molecular switches that reversibly transition between two states, M and AM, is illustrated by five trajectories (horizontal lines). Starting in state M, each switch stochastically (here with a time constant of 1 sec) reaches a thermally activated branch point where the switch either stays in state M or transitions to state AM. Here, for simplicity, I assume that $\Delta G° = 0$, which means that states M and AM are equally probable (a non-zero $\Delta G°$ simply changes the relative probability). The corresponding time course of the number of switches, $N_M(t)$, in state M (lower left) is determined statistically as a 50% decrease in the probability of being in state M at each branch (vertical dashed lines). The same exponential decay with a time constant of 1 second (curved line) is predicted from the time dependence of the number of ways, $W$, switches can be in state $N_M(t)$, $W(t) = N!/(N_M(t)!N_{AM}(t)!)$ (lower right), where here $N_M$ is a statistical expectation or microstate. This is essentially Boltzmann's equation applied to molecular switches, and an expression for $N_M(t)$ is formally developed elsewhere (17). These are two different statistical descriptions of the same irreversible chemical process, demonstrating a molecular-ensemble duality (18) that precludes predictive molecular models of irreversible ensemble processes. Molecular statistics describe the disordering of switches over time irreversibly pulled forward by an increase in entropy, $S = \ln W$, where here the number of possible ways, $W$, switches can exist at the $n^{th}$ branch point is $W = 2^n$, or $S = n{\cdot}\ln(2)$. Ensemble statistics describe a reaction energy landscape within which switches are pulled irreversibly with time toward an energy minimum with an increase in entropy, $S = \ln W = \ln[N!/(N_M!N_{AM}!)]$. Irreversible chemical reactions are described by ensemble statistics not by molecular statistics.

## 2.0 Muscle Physiology

Muscle is important for many physiological functions in humans. Smooth muscle is involved in the slow contractions of smooth organs like the stomach and intestines, moving food through our digestive system. Cardiac muscle makes our hearts beat, pumping blood through our circulatory system. And we engage skeletal muscle when we walk or lift things in moving our skeletal system. Muscle tissue consists of many muscle cells that contain long myofibrils composed of series of sarcomeres – the fundamental contractile unit of muscle. Within a sarcomere are interdigitated thin and thick filaments that slide past each other to make a sarcomere shorten. Many shortening sarcomeres arranged in series results in muscle contraction. Thin filaments are composed primarily of long actin protein polymers. Thick filaments are composed primarily of myosin molecules. The domain of a myosin molecule that creates a crossbridge between the actin thin and myosin thick filament undergoes a conformational change when it binds to the actin filament, pushing the actin filament. This is the mechanism of filament sliding. This actin-myosin binding step is repeated through the actin-myosin catalyzed ATP hydrolysis reaction, where the free energy for ATP hydrolysis is the energy available for generating force and power output by muscle.

### 3.0 Muscle Thermodynamics

In the early 1800's, S. Carnot set out to investigate the efficiency of a steam engine, and in 1824 he published Reflections on the Motive Power of Fire in which he established the foundation for thermodynamics, showing that the mechanics of individual molecules in the "working fluid" (steam) do not define the mechanical state of the engine (19). A century later, J. Langley suggested to A.V. Hill that he "investigate the efficiency of cut-out frog's muscle as a thermodynamic machine" (20). Hill went on to make precise measurements of work, $w$, and heat, $q$, output by shortening muscle, and in 1938 published his thermodynamic model of muscle contraction (2, 21). Like Carnot, Hill showed that the mechanics of individual molecules in muscle do not define the mechanical state of muscle. Specifically, Hill showed that the energy available for work by muscle, $\Delta G(F)$, decreases linearly with muscle force, $F$, which is to say that muscle mechanics and energetics are fully determined by the energetics and mechanics of the muscle system, not by the energetics and mechanics of individual molecules in that system (2).

A.V. Hill's equation (16, 17) simply describes the steady-state non-equilibrium thermodynamic transfer of energy, $\Delta G(F)$, to work, $w$, and heat, $q$, output, at a rate, $v$, or

$$v \cdot \Delta G(F) = dw/dt + dq/dt. \qquad \text{Eq. 1}$$

Hill developed this equation – the "framework into which detailed machinery must be fitted" (20) – knowing little or nothing about the detailed machinery of muscle.

### 4.0 Scaling: Molecule Versus Ensemble

Unfortunately, when the machinery of muscle began to be discovered, researchers did not attempt to fit this machinery into Hill's equation, rather they abandoned Hill's thermodynamics for molecular mechanic models. The rationale remains unclear. Hill had already demonstrated that thermodynamics applies to muscle, and Boltzmann had already demonstrated that the entropy of a thermodynamic system affects the energetics of molecules in that system. The assumption that macroscopic function can be described on any scale is simply false. For the same reason we cannot define an atomic model of muscle contraction, we cannot define a molecular model of muscle contraction. The reason is that the entropy of muscle's irreversible power stroke is defined only on the scale of muscle (Fig. 1).

### 4.1 The Scale of Myosin Molecules

The mechanics and chemistry of individual myosin motors have been directly measured in single molecule mechanics studies. Briefly, individual myosin motors are observed to function as force-generating switches induced by actin binding and gated by the release of inorganic phosphate, $P_i$. Here, a discrete chemical step (actin-myosin binding) is the chemical mechanism (22); the standard (molecular) free energy of actin-myosin binding, $\Delta G^\circ$, is the energy (9); a spring coarse-grained on the scale of a myosin motor is the mechanical element (6); and a discrete myosin lever arm rotation coarse-grained on the scale of a myosin motor that displaces the molecular spring a distance, $d$ (5, 22, 23), is the force-generating mechanism. Consistent with thermodynamics applied to an individual myosin motor, the force generated and the kinetics and energetics of force generation are fully defined on the scale of a myosin motor without defining the chemical forces of the thermally fluctuating structural components defined on smaller thermal scales (e.g., secondary and tertiary protein structures). This is because these thermally fluctuating structural components equilibrate on the scale of the myosin motor and define the entropy of the myosin motor on the scale of the myosin motor.

### 4.2 The Scale of Muscle

A.V. Hill showed that muscle functions on a larger thermal scale from that of individual myosin motors. The observed function is muscle force generation (1). A change in state of an ensemble of $N$ motors is the chemical mechanism (17); the reaction (system) free energy for actin-myosin binding, $\Delta G$, is the energy (16); a coarse-grained muscle spring is the mechanical element (17); and a coarse-grained ensemble lever arm rotation that displaces the coarse-grained muscle spring a distance, $d/N$ (17), is the force generating mechanism. The force generated and the kinetics and energetics of force generation are fully defined on the scale of muscle without defining the chemical forces of individual myosin motors. This is because these thermally fluctuating individual myosin motors equilibrate on the scale of muscle and define the entropy of muscle on the scale of muscle (Fig. 1). This is A.V. Hill's observation (2).

The above chemical mechanical descriptions are the same only they describe two different thermal scales. While molecular biophysicists readily accept the former, they do not to accept the latter, assuming instead that muscle mechanics and chemistry defined on the thermal scale of muscle can be described by myosin motor mechanics and chemistry defined on the thermal scale of individual myosin motors. The reason that a molecular power stroke mechanism is not possible is that the entropy, S, of an ensemble of myosin switches (i.e., a binary system), which increases ($\Delta S$) with muscle's irreversible power stroke, is not defined on the scale of individual myosin motors. Figure 1 illustrates how the reaction energy landscape of an ensemble of switches is defined by S only on the scale of the ensemble of switches.

We have shown that the molecular basis for A.V. Hill's simple thermodynamic model (Eq. 1) is that $v$ is the rate of the actin-myosin catalyzed ATP hydrolysis reaction, and $\Delta G(F) = \Delta H - T\Delta S - Fd/N$ is the free energy for actin-myosin binding, which energetically drives an ensemble of force-generating myosin switches to perform work against an ensemble force, $F$. Here, $\Delta H$ is the binding enthalpy (the standard free energy); $\Delta S$ is the reaction entropy of the binary system of switches (Fig. 1); $F$ is the force of the system spring; and $d/N$ is the displacement of that spring by a myosin switch.

The free energy for ATP hydrolysis energetically drives the entire reaction cycle. This includes detaching myosin motors from actin so that the binding energy, $\Delta G(F)$, can be used again for work upon myosin binding to actin. We have shown that this simple two-state model accurately accounts for most key aspects of muscle mechanics and energetics (14–17, 24).

### 5.0 Power Stroke

A power stroke is unambiguously defined as work performed over time by a smooth stroke. An oar stroke that moves a boat is a power stroke as is the stroke of a cilium that moves a protozoan. In swimming, a power stroke is required whereas in badminton and volleyball power strokes are not allowed. In a Carnot cycle, the isothermal expansion of a gas that moves a piston against an external force is a power stroke as is the shortening of muscle against an external force.

Discrete state chemical transitions and discrete mechanical switches occur faster than movement, and so while they can generate force, they are not power strokes because they do not generate movement over time. While to a non-scientist this might all seem like "splitting hairs", precise definitions of power strokes and force generation are necessary to accurately describe the mechanism of muscle contraction.

When muscle is held at a fixed length, myosin force-generating switches generate force in compliant elements in proteins throughout the muscle half sarcomere. When muscle shortens at a velocity, V, against a mechanical load, F, myosin force generating switches generate power output, FV. Hill's equation (Eq. 1) predicts a relationship between F and V from which an F-dependent efficiency can be derived (17, 25).

In general, muscle's power stroke occurs with the chemical relaxation of the actin-myosin binding reaction. For example, at a constant force, F, the ATP hydrolysis reaction perturbs the ensemble of myosin motors from a binding equilibrium and the subsequent chemical relaxation of the ensemble of myosin motors generates an ensemble power stroke. Any other perturbation to the actin-myosin binding equilibrium (e.g., a rapid change in F or a change in binding free energy, ΔG) can induce a power stroke through the subsequent relaxation of the binding reaction.

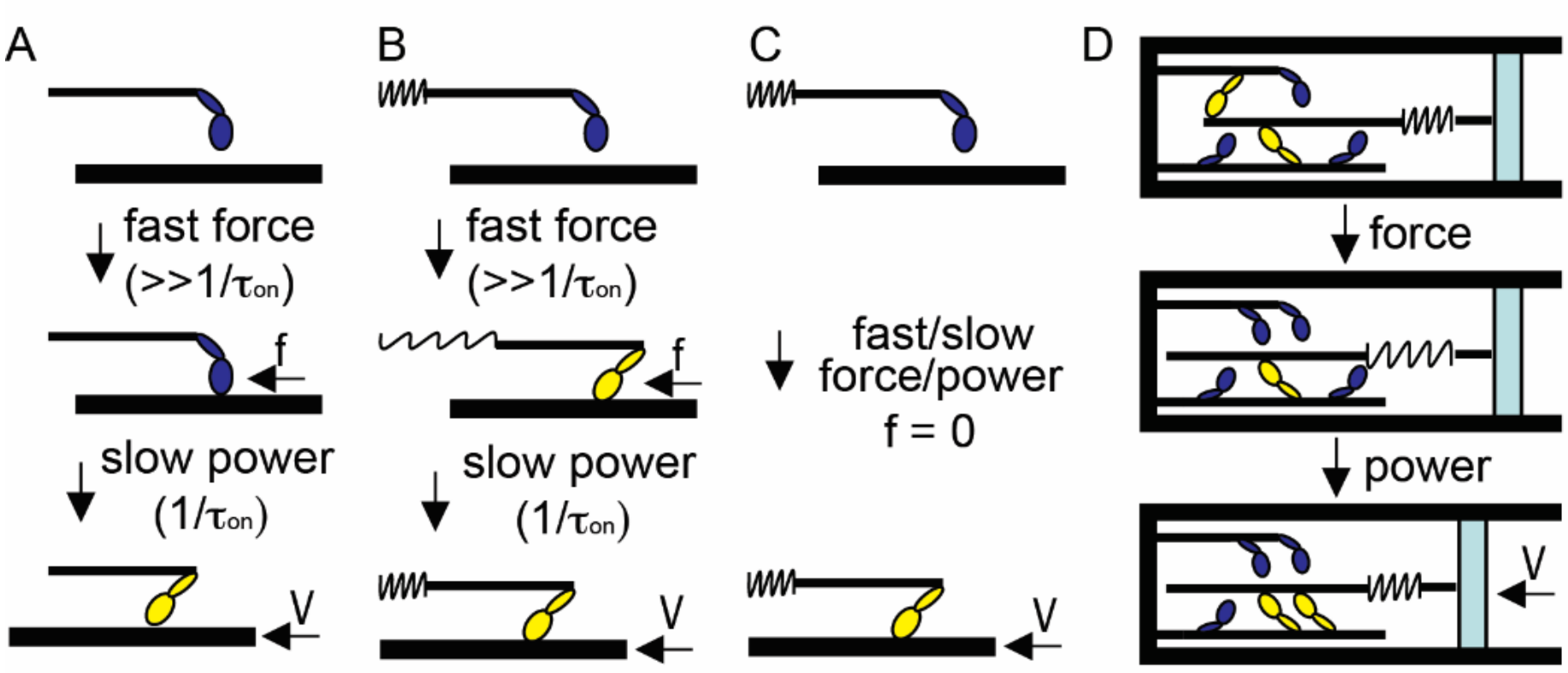

**Figure 2.** Four proposed power stroke mechanisms of muscle contraction. (A) According to a power stroke rotation model based on H.E. Huxley (26) and Lymn-Taylor (27), an individual myosin motor (ovals) generates force, *f* (top to middle) upon binding an actin filament (black rectangle) and subsequently generates a power stroke when a smooth myosin rotation moves actin, *V* (middle to bottom). (B) According to a force-generating rotation model based on Huxley-Simmons(28), an individual myosin motor rotation induced by actin binding generates force, *f*, by stretching a compliant element in that motor (top to middle) and subsequently generates a power stroke when the shortening molecular spring moves actin, *V* (middle to bottom). (C) A force-generating molecular power stroke generates both force (stretches a spring) and power (shortens a spring) through a myosin motor rotation induced by actin binding, which is to say it generates neither force, *f* = 0, nor power. (D) Thermodynamic force generation is not localized to a molecular spring; it is collectively generated in a system spring by an ensemble of myosin motor switches. Force generation in the system spring along the actin-myosin binding isotherm shifts the equilibrium distribution of switches toward unbound myosin heads (top to middle) and the subsequent shortening of the system spring along the isotherm (a thermodynamic power stroke) shifts the equilibrium distribution of switches back toward bound myosin heads (middle to bottom).

### 5.1 A.F. Huxley's Molecular Power Stroke

In 1662, based on his corpuscularian philosophy, Robert Boyle proposed that the molecular mechanism for gas pressure was compressed springs of air (29). That is, Boyle assumed that force defined on the thermal scale of a cylinder is described by forces defined on the thermal scale of individual gas molecules. Carnot disproved Boyle's corpuscular mechanic "springs of air" mechanism for gas pressure, showing that the pressure of a gas is not determined by the mechanics of individual gas molecules. Similarly, in 1938 A.V. Hill disproved a corpuscular mechanic "springs of myosin" mechanism for muscle force in which stretched springs of myosin motors determine muscle force. Unfortunately, corpuscular mechanics poisons the minds of even the greatest scientists, and in 1957, models of muscle contraction began an unprecedented scientific regression from Carnot's $19^{th}$ century thermodynamics to Boyle's $17^{th}$ century corpuscularianism to a structural determinism invoked today that predates science (see below).

In 1957 A.F. Huxley observed individual myosin crossbridges (or motors) within the macromolecular protein assembly of a muscle sarcomere (3), and he reasoned that if individual myosin motors are observable in muscle then muscle mechanics must be observable within individual myosin motors (6). Based on this corpuscular mechanic reasoning, he developed a springs of myosin model, challenging thermodynamics and A.V. Hill's observations.

Huxley redefined the power stroke mechanism of muscle contraction from A.V. Hill's thermodynamic power stroke energetically driven by $\Delta G(F)$ to a molecular power stroke (Fig. 2A or Fig. 2B) energetically driven by the shortening of a spring of myosin. Because corpuscular mechanics had never before been formalized, a new molecular chemistry was needed to replace Gibbs' ensemble chemical thermodynamics (30). This new molecular chemistry allowed for the force-dependence of the energetics of individual molecules to be arbitrarily defined giving Huxley's model the flexibility to describe most any force-dependent power output relationship. In contrast, the linear force-dependent energetics observed by A.V. Hill directly accounts for the force-dependent power output observed in muscle.

The difference between Gibbs' ensemble chemistry and molecular chemistry is obvious. The Gibb's reaction free energy describes the change in free energy of an ensemble of molecules as a function of ensemble entropy:

$$\Delta G(F) = \Delta H - T\Delta S,$$

whereas molecular chemistry describes a change in the "basic free energy", $\mu_i$, of an individual molecule, i

$$\Delta\mu_I = \Delta H,$$

independent of ensemble entropy, where here H is a molecular enthalpy.

One can only speculate how, by any measure, Huxley's corpuscular mechanic model prevailed over thermodynamics despite well-founded objections from A.V. Hill and from my research group over the past quarter century (14). Perhaps it was a lack of sufficient training in thermodynamics and statistical mechanics or perhaps it was aspects of thermodynamics (in particular entropy) (11) that were underdeveloped. As shown below, it is clear that corpuscularianism remains deeply rooted in the thought processes of many modern scientists, and twenty-five years of reviewer comments document repeated forceful rejections of thermodynamics in defense of $17^{th}$ century corpuscularianism, revealing corpuscularianism to be a deeply personal, emotional, and irrational conviction. Whatever the reason, corpuscularianism prevailed and led an entire scientific community on a 65-year quest to disprove thermodynamics in search of the springs of myosin that shorten with muscle's power stroke.

#### 5.1.1 The 65-Year Search for a Molecular Power Stroke

Prior to 1970, the phrase "power stroke" was primarily used in the biological sciences to describe the macroscopic power strokes of muscle, cilia, etc. Not until the late 1960's did x-ray diffraction and birefringence experiments (26, 31, 32) on muscle fibers begin to pique interest in the specific structural change in a myosin motor associated with Huxley's molecular power stroke mechanism. Two different mechanisms were soon proposed.

#### 5.1.2 H.E. Huxley and the Huxley-Simmons Molecular Power Strokes

H.E. Huxley's 1969 power stroke (Fig. 2A) is a smooth rotation of a tensioned myosin motor (26) that occurs with muscle shortening. In 1971, Lymn and Taylor coupled this power stroke mechanism to the actin-myosin ATPase reaction cycle (27) in a model that describes force-generation (tensioning a myosin spring) upon actin-myosin binding (Fig. 2A, top to middle), and a subsequent power stroke rotation (the shortening of the spring) that moves the actin filament a distance, *d*, over the lifetime, $\tau_{on}$, of the actin-bound state at a rate, $V = d/\tau_{on}$ (Fig. 2A, middle to bottom). This is the power stroke mechanism illustrated in most textbooks today. In contrast, in the Huxley-Simmons 1971 model (Fig. 2B), a force-generating myosin rotation (28), or switch, occurs much faster than muscle shortening, rapidly stretching a molecular spring (Fig. 2B, top to middle) that subsequently shortens with a power stroke (Fig. 2B, middle to bottom). Both models can be described by T.L. Hill's corpuscular mechanic formalism (30), which is the foundation of every molecular model of muscle contraction to date. In H.E. Huxley's model (Fig. 2A), the rotation from a pre- to post-power stroke state is continuous; it occurs with a decrease in force; and it requires muscle shortening. In the Huxley-Simmons model (Fig. 2B), the rotation from the pre- to post-switch state is discrete; it generates force; and it occurs on a time scale much faster than muscle shortening.

According to a Google Scholar search (Fig. 3), results for "muscle myosin 'power stroke'" first appear in 1972 (33–35), and throughout the 1970's a myosin power stroke was described exclusively as the textbook myosin structural change (Fig. 2A) occurring with muscle shortening. However, in the late 1970's and early 1980's researchers began to observe two distinct structural states of myosin in muscle indicative of a switch (36, 37). In 1982 the D.D. Thomas and R. Cooke groups observed in active isometric muscle that all actin-bound myosin motors were, according to the textbook model, in a post-power stroke state (38) counter to the expectation that in a non-shortening muscle actin-bound myosin motors would be in a pre-power stroke state. This observed actin-bound myosin structure was clearly in a post-switch not a post-power stroke state indicative of a force-generating switch induced by actin binding (Fig. 2B). However, rather than conclude that the textbook molecular power stroke mechanism was wrong, in the early 1980's a force-generating switch induced by actin binding was referred to as a force-generating power stroke mechanism (36, 38), where "power stroke" was added to the name based solely on a conviction that a molecular power stroke existed.

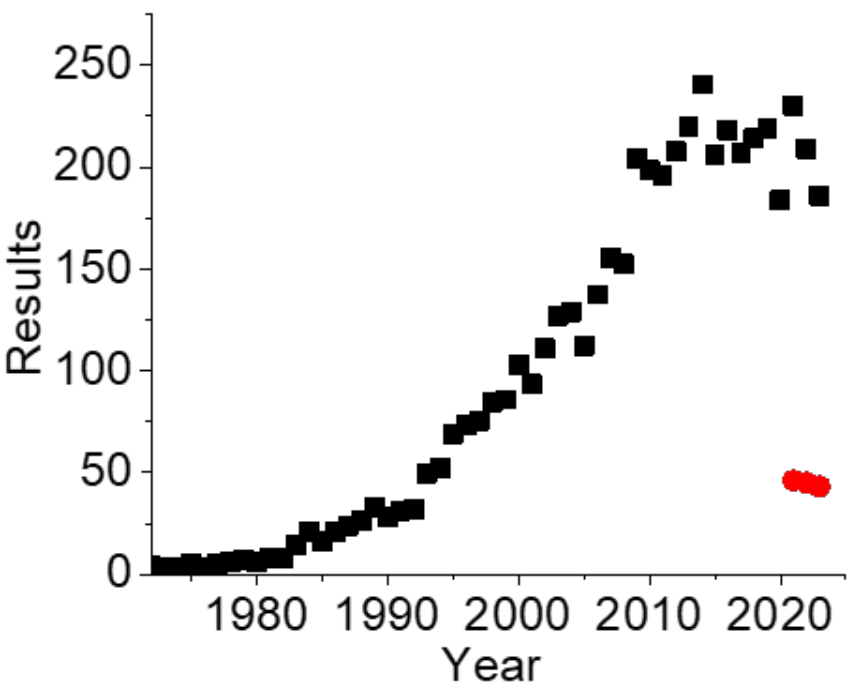

**Figure 3.** Results of a Google Scholar search for "muscle myosin 'power stroke'". Results that do not describe one of the power stroke mechanisms in Fig. 2 or results where the mechanism described was unclear were excluded from this analysis. The remaining number of results filtered by year from 1972 to 2023 are plotted (black squares). For 2021, 2022, and 2023 the number of search results that describe a power stroke as a myosin structural change that occurs with muscle shortening are plotted (red circles).

### 5.1.3 The Paradoxical Molecular Force-Generating Power Stroke

A "force-generating power stroke" mechanism is impossible to model formally (Fig. 2C) because a discrete structural change that both stretches a spring and shortens a spring at the same time does neither. Thus, the force-generating power stroke mechanism is a paradox not a mechanism, posing the question: how can a single force-generating switch account for two completely different mechanisms of muscle contraction (mechanical energy stored in muscle and mechanical energy lost from muscle) that occur on two completely different time scales (bond formation vs. muscle shortening)?

The belief in a power stroke mechanism was so great that "models" transitioned from physical to conceptual power stroke mechanisms where unphysical and unobserved intermediate states of a binding-induced switch were introduced to make binding less discrete. For example, a weak-to-strong binding mechanism described actin-myosin binding as a relatively smooth process that occurs when a myosin motor "rolls" along an actin filament from weak toward strong binding states (39), turning a binding induced force-generating switch into more of a graded force-generating power stroke. These invented mechanisms were, however, eventually disproven and the final defense of the molecular power stroke paradigm became nothing more than word play.

In 1987, H.E. Huxley wrote "I have always felt somewhat uneasy with the concept of the myosin head undergoing a very large and progressive shape change [a power stroke, Fig. 2A], and much prefer the picture of two well defined structural states with a very rapid transition between them [a force generating switch, Fig. 2B]" (40). He then proposed a force-generating switch mechanism resembling that of Huxley-Simmons and referred to it as a power stroke (40). In 1993, I. Rayment published a high-resolution crystal structure of myosin subfragment-1, indicating that the myosin rotation occurs with the rotation of a myosin lever arm (an alpha helix with two bound light chains) (41). In describing this important discovery, E. Taylor highlighted what remained to be determined: "Part of the energy in ATP hydrolysis is stored mechanically as a 'stretched spring' in the head or alternatively in the connection of the head to the myosin filament" (42), the former describing the lever arm rotation as a power stroke (Fig. 2A) and the latter describing the lever arm rotation as a force-generating switch (Fig. 2B). Even though there was mounting evidence for the latter, a power stroke would from here on be defined structurally, not mechanically, as a lever arm rotation. A dramatic increase in "power stroke" references followed (Fig. 3), and in 1997 even A.F. Huxley referred to his force-generating switch as a power stroke (43).

And so it was, the force-generating myosin motor switch – paradoxically named a force generating "power stroke" because a molecular power stroke did not exist – had now become the molecular "power stroke" mechanism. Corpuscularians-now-turned-structural-determinists declared victory and today confidently claim that there is overwhelming evidence that the mechanism of muscle's power stroke is a molecular "power stroke" that is not a power stroke.

**5.2 Muscle's Thermodynamic Power Stroke**

In 1998, we observed in muscle a discrete force-generating myosin lever arm rotation induced by actin binding and gated by phosphate release with no evidence for a molecular power stroke (7). Our kinetic scheme – the first to illustrate an actin-induced switch – was widely rebuffed on the basis that actin binding must precede the "power stroke" (44). Single molecule mechanics studies similarly provided direct evidence for a force-generating myosin switch induced by actin binding and gated by inorganic phosphate release with no evidence for a molecular power stroke (5, 22). Single molecule structural studies similarly provided direct evidence for a force-generating myosin switch induced by actin binding and gated by inorganic phosphate release with no evidence for a molecular power stroke (45).

It was clear that the myosin lever arm rotation was a force-generating switch (Fig. 2B). It was also clear that the same force-generating switch, with all previously proposed molecular power stroke mechanisms now disproven, was the only feasible molecular mechanism for muscle's power stroke. Thus, the force-generating power stroke paradox – a version of Loschmidt's paradox – remained: how do reversible force-generating switches generate irreversible muscle power output?

A solution to this paradox came when force and the ensemble state of switches were simultaneously measured in muscle (7, 9). These experiments showed that the free energy for actin myosin binding (i.e., the force-generating switch) is a function of muscle force, $F$, and is the energy available for collective force generation and power output (9)

– precisely what A.V. Hill showed in 1938 (2, 16). These results implied that muscle force generation and power output are two different thermodynamic mechanisms that both have the same molecular mechanism (a force-generating switch) (17). I first presented this model in 1998 as a graduate student in a conference talk titled "muscle contracts with the flip of an ensemble of switches" (46). When I stated that muscle's power stroke is analogous to the power stroke in a Carnot cycle, the audience burst into laughter.

In 2000, we abandoned A.F. Huxley's assumption that muscle structure and function are coarse-grained on the thermal scale of a myosin motor and considered instead that muscle structure and function are coarse-grained on the thermal scale of muscle. That is, we defined muscle mechanics in terms of a muscle – not a molecular – spring, where a net flux of force-generating switches through the actin-myosin ATPase reaction cycle generates force collectively in that spring. We similarly defined kinetics and energetics in terms of ensemble (Gibbs) – not molecular (T.L. Hill) – chemistry (16, 17).

### 5.2.1 The Entropic Power Stroke Mechanism Reconciles Gibbs and Boltzmann Entropy

The difference between the free energy of a myosin motor (standard free energy) and the free energy of an ensemble of myosin motors (reaction free energy) is the entropy of a binary system of mechanical switches in the ensemble chemical state, {bound, unbound}. For example, there are fewer ways ($W = 5!/(4!1!) = 5$) that one out of five switches can be bound to actin in state {1,4} (Fig. 2D, top) than there are ways ($W = 5!/(3!2!) = 10$) that two out of five switches can be bound to actin in state {2,3} (Fig. 2D, middle). Because the entropy, $S = k_B \cdot \ln W$, of the former is less than the entropy of the latter, the reaction is energetically and kinetically pulled toward binding (toward Fig. 2D, top) (17, 47). This is Boltzmann's entropy, and it is related to Gibbs' reaction entropy, $\Delta S$, as follows.

According to Gibbs, with a change in chemical state from {1,4} to {2,3}, the change in entropy is $\Delta S = k_B \cdot \ln(10/5) = k_B \cdot \ln 2$, and so $-T\Delta S = -k_B T \cdot \ln 2$. In addition to the standard (molecular) free energy, $-T\Delta S$ contributes to the reaction (system) free energy for this change in state. In general, for a change in chemical state from i to i + 1, $T\Delta S = k_B \cdot \ln(\Omega_{i+1}/\Omega_i)$, which I have shown mathematically resembles the term in the reaction free energy often described as a change in chemical activity, $\mu_{i+1} - \mu_i$, implying that while chemical activity describes the ensemble free energy change when molecules are added to the system, entropy (not chemical activity) describes the ensemble free energy change when a fixed number, $N$, of molecules in a system change chemical state (17, 47).

According to the Gibbs reaction free energy equation, the relationship between $F$ and $T\Delta S$ is that of an entropic spring (17), where an increased system force energetically pulls the system toward lower entropy with fewer myosin switches bound to actin (47). Through this entropic mechanism, muscle force, $F$, defines the chemistry of myosin motor switches, which means, as shown by A.V. Hill, it is physically impossible for the chemistry of myosin motor switches to define muscle force. The molecular details in Fig. 2D are purely illustrative because molecular geometries and mechanics are all delocalized through thermal fluctuations on the time scale of muscle contraction. The chemical states illustrated in Fig. 2D represent thermally equilibrated microstates, only one of which is depicted.

A thermodynamic power stroke is defined along a binding isotherm (17) at a force that is reached through collective force generation in the system spring (a net flux of force-generating switches through the ATPase cycle) (13). Figure 2D (top to middle) illustrates how when the system spring is isothermally stretched along the isotherm by a net flux of force-generating switches (13) (not shown), the binding equilibrium shifts toward unbound motors at higher forces. A decrease in bound motors with an increase in force (17) is inconsistent with corpuscular mechanics but is consistent with Le Chatelier's principle in thermodynamics. The thermodynamic power stroke (Fig. 2D, middle to bottom) occurs with isothermal shortening of the system spring which shifts the binding equilibrium back toward motor attachment.

Like in a Carnot cycle, entropy increases with a power stroke and is conserved around the thermodynamic work loop of muscle (the entropy gained through a power stroke is lost upon force generation, Fig. 2D). However, here the chemistry underlying this conserved entropy is clear. The entropy is that of a binary system, and an increase in entropy energetically drives actin binding (Fig. 2D), defining the irreversible energetics and kinetics (the arrow of time) of muscle's thermodynamic power stroke (47).

#### 5.2.2 The Entropy of a Binary Mechanical System is Created by Thermal Scaling

The entropy of a binary mechanical system of *N* myosin motor switches does not exist in *N* individual switches thermally isolated in separate containers (e.g., in *N* separate single molecule mechanics experiments). This ensemble entropy is physically created when *N* individual switches become part of a single binary system containing *N* switches. This increase in entropy occurs when thermal energy becomes delocalized among the N switches in the form of heat, creating a resonant ensemble structure that is energetically and irreversibly stabilized by this increase in entropy (12). This is thermal scaling, and it is a physical process not a mental exercise. It is physically observed in muscle (9); it physically changes the kinetics and energetics of *N* myosin motor switches (47); and it physically accounts for the irreversible kinetics and energetics of muscle contraction (17). With thermal scaling, entropy, heat and free energy are all quantized by the number of molecules, *N*, in an ensemble, leading to quantum-like observational uncertainty across thermal scales and entanglement within thermal scales.

The basic mechanism for entropically driven thermal scaling is simple yet profound. While we often associate an increase in entropy with disorder, here entropy creates higher-order structures because, in a given state at a given instant of time, more microstates exist on the thermal scale of the ensemble than exist on the thermal scale of individual molecules. This implies that in addition to structure, energetics, kinetics, heat, and entropy, the definition of an instant of time physically changes across thermal scales. This is consistent with ensemble entropy creating the irreversible kinetics of the ensemble, increasing the length of the arrow of time with thermal scaling.

#### 5.2.3 A Confluence of Physics and Biology: Top-Down Meets Bottom-Up

It is no surprise that muscle – a thermodynamic system in which chemistry, mechanics, and structure are well-defined on two different thermal scales – provided the vehicle for establishing thermal scaling at the confluence of physics and biology. Specifically, from the penultimate to the actual thermal scale of muscle, *N* individual force-generating myosin motor switches collapse into an ensemble of *N* motor switches creating the entropy of a binary system that defines muscle's resonant ensemble structure and the irreversible kinetics and energetics of muscle's power stroke.

The typical bottom-up approach of biologists focuses on classical mechanic descriptions of how atoms scale up through primary, secondary, tertiary, quaternary, and macromolecular structures to create functional biological systems. However, this approach will never converge with muscle thermodynamics because muscle chemistry and mechanics are defined independent of the component structures of muscle at every thermal scale below that of muscle. Biological structures and their energetics are certainly created bottom-up; however with thermal scaling, structural components physically collapse into resonant ensemble structures creating entropy that defines both the resonance (delocalized heat) and the irreversible kinetics and energetics (a change in that entropy) of functional biological structures. In other words, biological structures and their energetics are physically created from the loss of identity of their structural components and thus cannot be described classically in terms of those structural components. In this way, thermal scaling resembles quantum mechanics only on the scale of thermal energy (12).

## 6.0 Corpuscular Mechanics is Insidious

A thermodynamic power stroke resolves the force-generating power stroke paradox; it accurately describes most key aspects of muscle contraction (13, 15–17); and it has significant implications for physical chemistry (11–13, 17). Nevertheless, for the past 25 years an entire scientific community has not only ignored the thermodynamic solution to the force-generating power stroke paradox; they have ignored the paradox, simply declaring that the force-generating switch is a molecular "power stroke" (Fig. 3). That is, this community collectively chose to abandon all physical reason rather than conclude that A.F. Huxley's molecular power stroke mechanism – a mechanism that violates the laws of thermodynamics and ignores A.V. Hill's observations – was wrong. As a result, this community today invokes structural determinism across mixed thermal scales (48, 49), where descriptions of structures (e.g., a lever arm rotation) supersede descriptions of muscle function (e.g., a power stroke).

Naming a force-generating switch a "power stroke" is not an inconsequential misnomer. Figure 3 shows that over the past three years < 25% of the search results for "muscle myosin 'power stroke'" (Fig. 3, red circles) describe a textbook molecular power stroke mechanism (Fig. 2A). Of these, more than 90% are student theses, book chapters, and physiology, kinesiology, clinical, theoretical, and educational papers. The authors of the other >75% of papers are experts who know that the myosin rotation is a force-generating switch and thus deceive and confound the broader scientific community by referring to this switch as a power stroke. Declaring that a force-generating molecular switch is a molecular "power stroke" is the double speak of conspiracy theorists, hiding the fact that after 65 years of rejecting muscle thermodynamics in search of the springs of myosin, an obsolete 17th century philosophy has failed to answer the fundamental question: what is the power stroke mechanism of muscle contraction?

## 7.0 Discussion

There is overwhelming evidence that the molecular mechanism of muscle contraction is a force-generating switch induced by actin binding and gated by inorganic phosphate release. While the molecular power stroke model proposed by H.E. Huxley in 1969 was the first to involve a myosin rotation (Fig. 2A), his proposed rotation occurs on a time scale, $\tau_{on}$, that is much slower than the time scale of bond formation. While the Huxley-Simmons force-generating switch (Fig. 2B) resembles that observed experimentally, this switch generates force in a molecular spring that shortens with a subsequent molecular power stroke. There is no evidence for this molecular spring. A myosin switch (the lever arm) is a relatively rigid alpha helix stabilized by two light chains that generates force in compliant elements external to the switch (9, 17, 22) (i.e., in a system spring), which is to say the subsequent power stroke is not molecular. A "force-generating power stroke" (Fig. 2C) describes not a mechanism but a paradox of how one molecular mechanism (a force-generating switch) can be the mechanism for both muscle force generation on the time scale of bond formation and muscle power output on the time scale of muscle shortening. A thermodynamic power stroke (Fig. 2D) solves this paradox, describing muscle power output as the shortening of an entropic spring containing an ensemble of force-generating myosin switches.

That Carnot in 1824 and A.V. Hill in 1938 disproved corpuscular mechanics should not offend scientists. That scientists have for over 65 years rejected and ignored Carnot and Hill should. Corpuscularianism infects the minds of even the greatest scientists. Boyle's corpuscular mechanic "springs of air" model of gas pressure is wrong, but that has not diminished his contributions to chemistry, the scientific method, or Boyle's law. Huxley's corpuscular mechanic "springs of myosin" model of muscle force is equally wrong, but that does not diminish his contributions to physiology (28), the sliding filament theory (3), or Hodgkin-Huxley (50). However, when we fail to acknowledge and vigilantly address this common flaw in the thought process of scientists, corpuscularianism spreads (< 0.5% of the search results in Fig. 3 describe a thermodynamic power stroke), which is a grave disservice to science and the public.

No doubt, the desire is strong to determine mechanisms of biological function at increasingly smaller thermal scales, but thermodynamics physically precludes this. Carnot's demonstration that on account of entropy the mechanics of molecules within a system do not determine the mechanical state of that system applies at every thermal scale in biological systems (12, 47). Muscle thermodynamics shows that mechanisms of biological function are only defined by the entropically-stabilized coarse-grained structure, mechanics, and chemistry physically defined on the scale of biological function (12).

A thermodynamic muscle model has a limited number of parameters (e.g., the number of motors, the binding free energy, and so testing the model and using the model to modulate muscle performance should focus on these parameters. For example, the model predicts that muscle force increases linearly with the binding free energy. Therefore, the model can be tested by correlating binding free energy and muscle force, and small molecules that enhance binding free energy would be expected to enhance muscle force. We have established many model predictions that can be experimentally tested.

**Table 1. Glossary of Terms**

| **Term** | **Definition** |
|---|---|
| Muscle Force, *F* | The end-to-end force of muscle generated within Compliant elements of the sarcomere |
| Muscle Power Stroke | Work performed by shortening muscle over time against a load, *F*. |
| Muscle Shortening Velocity, *V* | The rate at which a muscle half sarcomere shortens. |
| Work, w | The product of the distance a muscle shortens and the force against which it shortens. |
| Heat, *q* | The measured heat output by shortening muscle. |
| Power, *FV* | Work performed over time. The product of the muscle shortening velocity and the force against which it shortens. |
| Boltzmann's Entropy | $k_B\ln(W)$, where *W* is the number of ways molecular switches can exist in an ensemble state. In the state where $N_M$ switches are in state M and $N_{AM}$ switches are in state AM, $k_B\ln[N!/(N_M!N_{AM}!)]$ |
| Gibb's Entropy, ΔS | The change in Boltzmann's entropy with a change in the state of an ensemble of switches. |
| Myosin | The contractile protein of the thick filament |
| Myosin Motor | The part of myosin that creates a crossbridge between the thick and thin filaments. The myosin motor binds and catalyzes the hydrolysis of ATP and binds actin. It is also called the myosin head. |
| Actin | Proteins that form the polymeric thin filament. |

## 8.0 Conclusion

The fascinating history of this 65-year search for the springs of myosin in defiant opposition to a well-developed (2, 16, 17) and tested (13–17) thermodynamic muscle model is beyond the scope of one article. However, the lesson moving forward is clear. Biological function evolved constrained by the laws of thermodynamics, which means that we will never understand how biological systems work without first understanding Carnot's arguments published 200 years ago this year.

**Funding:** This work was funded by a grant from the National Institutes of Health 1R01HL090938-01.

**Data Availability Statement:** No new data were created.

**Acknowledgments:** I thank Julie and my students and mentors over the past 25 years for their insights and inspiration.

**Conflicts of Interest:** The author declares no conflicts of interest.